\documentclass[twocolumn,showpacs]{revtex4}
\usepackage{amsmath,amsthm,amssymb,amscd,float,accents}
\usepackage{graphicx}
\usepackage[T1]{fontenc}
\usepackage{lmodern}
\newcommand{\al}{\alpha}
\newcommand{\be}{\beta}
\newcommand{\de}{\delta}
\newcommand{\ep}{\epsilon}

\newcommand{\si}{\sigma}
\newcommand{\te}{\theta}
\newcommand{\vp}{\varphi}

\newcommand{\ZZ}{{\mathbb Z}}
\newcommand{\cH}{{\mathcal H}}

\newcommand{\pa}{\partial}

\let\ni\noindent
\newcommand{\ms}{\mspace{1mu}}

\renewcommand{\le}{\leqslant}

\def\bbuildrel#1_#2^#3{\mathrel{\mathop{\kern0pt #1}\limits_{#2}^{#3}}}
\newcommand{\tends}[1]{\bbuildrel{\hbox to 2em{\rightarrowfill}}_{#1}^{}}
\newcommand{\erf}{\operatorname{erf}}

\newcommand{\iu}{\mathrm{i}}
\newcommand{\e}{\mathrm{e}}

\newcommand{\diff}{\mathrm{d}}

\newcounter{ex}
\def\cond{\stepcounter{ex}\hskip-.75cm
  \makebox[.55cm][r]{(\roman{ex})}\hskip.2cm}

\begin{document}
\title{$1/f^\al$ noise and integrable systems}
\author{J. C. \surname{Barba}}
\author{F. \surname{Finkel}}
\author{A. \surname{Gonz\'alez-L\'opez}}%
\author{M. A. \surname{Rodr\'\i guez}}
\affiliation{Departamento de F\'\i sica Te\'orica II, Universidad Complutense, 28040 Madrid, Spain}
\date{May 14, 2009}
\begin{abstract}
  An innovative test for detecting quantum chaos based on the analysis of the spectral
  fluctuations regarded as a time series has been recently proposed. According to this test,
  the fluctuations of a fully chaotic system should exhibit $1/f$ noise, whereas for an
  integrable system this noise should obey the $1/f^2$ power law. In this letter, we
  show that there is a family of well-known integrable systems, namely spin chains of
  Haldane--Shastry type, whose spectral fluctuations decay instead as $1/f^4$. We present
  a simple theoretical justification of this fact, and propose an alternative characterization
  of quantum chaos versus integrability formulated directly in terms of the power spectrum of the
  spacings of the unfolded spectrum.
\end{abstract}
\maketitle
%

In the absence of a formal definition of quantum chaos, several different tools have been used in
the literature to distinguish the chaotic versus integrable character of a quantum system. The
most widely used among them is the study of the density of normalized spacings between consecutive
levels of the ``unfolded'' spectrum, which gives a quantitative estimate of the local level
fluctuations. More precisely, for quantum systems whose corresponding classical limit is chaotic,
the celebrated Bohigas--Giannoni--Schmit conjecture~\cite{BGS84} posits that this density should
coincide with that of a suitable Gaussian ensemble in random matrix theory (RMT). On the other
hand, a long-standing conjecture of Berry and Tabor~\cite{BT77} states that the spacings
distribution of a ``generic'' integrable quantum system should follow Poisson's law. Both of these
conjectures have been verified for a large number of systems, both numerically and using
semiclassical analytic methods~\cite{GMW98-4}.

More recently, an alternative characterization of quantum chaos was proposed in an interesting
paper by Rela\~no et al.~\cite{RGMRF02}. The basic idea behind this characterization is considering
the sequence of energy levels as a time series, whose corresponding power spectrum is then
analyzed using standard procedures. It was observed in the latter paper that for all three
classical random matrix ensembles the (averaged) power spectrum of the fluctuations of the
spacings exhibits $1/f$ noise, while the noise of a spectrum with a Poissonian spacings
distribution behaves as $1/f^2$ (the so-called Brown noise). These numerical observations were
theoretically explained using the machinery of RMT in Ref.~\cite{FGMMRR04}, where it was also
conjectured that the $1/f$ (resp.~$1/f^2$) noise of the power spectrum is in fact a universal
property of \emph{all} fully chaotic (resp.~integrable) quantum systems. Further numerical
simulations have lent additional support to this conjecture. More precisely, the $1/f$ law has
also been detected in the two-body random ensemble~\cite{RMR04}, whereas a more general power law
$1/f^\al$ with $1\le\al\le2$ is observed for quantum billiards (or potentials) and random matrix
ensembles interpolating between fully chaotic and integrable regimes, with $\al$ respectively
attaining the values $1$ and $2$ in these limits \cite{GRRFSVR05-4}.

An essential ingredient in the theoretical justification of the $1/f^2$ law for the power spectrum
of integrable systems is the Poissonian character of the spacings distribution~\cite{FGMMRR04}.
However, in a series of recent papers~\cite{FG05,BB06,BFGR08,BFGR08epl,BFGR09-3} it has been shown
that there a is wide class of integrable quantum models whose spacings distribution is not
Poissonian, namely spin chains of Haldane--Shastry (HS) type~\cite{Ha88-3}. These models, which
are the prime examples of integrable spin chains with long-range interactions, appear in
connection with several phenomena of physical interest such as strongly correlated
systems~\cite{ASK01}, generalized exclusion statistics~\cite{MS94-2}, and the AdS-CFT
correspondence~\cite{HL04}. It is therefore natural to ascertain whether the $1/f^2$ law
conjectured in Ref.~\cite{FGMMRR04} applies to spin chains of HS type. These chains are
particularly good candidates for a time series analysis, since their spectrum can be exactly
computed for a very large number of sites, and there is ample numerical and theoretical evidence
that their level density (whose knowledge is essential for unfolding the spectrum) becomes
Gaussian when the number of sites is sufficiently
large~\cite{EFGR05,FG05,BB06,BFGR08,BFGR08epl,BFGR09-3,EFG09}. In this letter we shall focus on
the simplest type of HS chains, whose raw spectrum is equally spaced (as in the Polychronakos
chain) or almost equally spaced (as in the original Haldane--Shastry chain). Our main result is
that the power spectrum of these chains ---unlike all the integrable models previously studied in
the literature--- clearly obeys the $1/f^4$ power law, characteristic of black noise. In order to
highlight the main ideas and simplify the theoretical derivation we shall primarily concentrate on
the case of an equispaced raw spectrum, giving only a qualitative justification of our result in
the general case.

Let $E_1<\cdots<E_{n+1}$ be a spectrum of a quantum system, and denote by $\mu$ and
$\si$ its mean and standard deviation. In accordance with the previous remark, we shall assume that
the following conditions are satisfied:

{\leftskip.75cm\parindent=0pt%
\cond The energies are equispaced, i.e., $E_{j+1}=E_1+jd$.

\cond The continuous part of the level density is a Gaussian with parameters
$\mu$ and $\sigma$, with $\si\gg\sqrt n\,d$.

} \ni The requirement that $\si\gg\sqrt n\,d$, which shall be needed in what follows, is
satisfied by all spin chains of HS type studied so far when the number of sites is sufficiently
large~\cite{BFGR08,BFGR08epl,BFGR09-3}. By the second condition, the unfolded spectrum is
given by $\ep_i=F(E_i)$, where
\[
  F(E)=\frac1{\sqrt{2\pi}\si}\int_{-\infty}^E\e^{-\frac{(x-\mu)^2}{2\si^2}}\diff x
  =\frac12\bigg[1+\erf\bigg(\frac{E-\mu}{\sqrt 2\,\si}\bigg)\bigg].
\]
The $i$-th normalized spacing of the unfolded spectrum is then defined by
$s_i=n(\ep_{i+1}-\ep_i)/(\ep_{n+1}-\ep_1)$, so that the mean spacing is one.
Following~\cite{RGMRF02}, we shall characterize the fluctuations of the spacings from their mean
by the statistic $\de_l = \sum_{j=1}^l(s_j-1)$, $l=1,\dots,n$, whose discrete Fourier transform is
given by
\begin{equation}\label{hdek}
\hat\de_k=\frac1{\sqrt n}\,\sum_{l=1}^n\de_l\,\e^{-\frac{2\pi\iu\ms k l}n}\,,\qquad k=1,\dots,n\,.
\end{equation}
In this letter we shall be interested in the behavior of the power spectrum of the statistic
$\de_l$, defined by
\[
P(k)=|\hat\de_k|^2\,,\qquad k=1,\dots,n\,.
\]
Note that, since the $\de_l$'s are real, $\hat\de_{n-k}$ is the complex conjugate of $\hat\de_k$, so
that $P(n-k)=P(k)$. For this reason, from now on we shall take $k$ in the range $1,2,\dots,[n/2]$.

To begin with, we shall prove a remarkable identity expressing $P(k)$ directly in terms
of the Fourier transform of the spacings $s_j$. Indeed, taking into
account that
\[
\sum_{l=1}^nl\,\e^{-\frac{2\pi\iu\ms k l}n}=-\frac{\pa}{\pa t}\bigg|_{t=\frac{2\pi\iu k}n}
\sum_{l=1}^n\e^{-tl}
=-\frac{n}{2\iu}\,\frac{\e^{-\frac{k\pi\iu}n}}{\sin(\frac{k\pi}n)}\,,
\]
from Eq.~\eqref{hdek} we obtain
\begin{equation}
  \label{deksj}
  \hat\de_k = \frac1{\sqrt n}\sum_{l=1}^n\sum_{j=1}^l s_j\,\e^{-\frac{2\pi\iu\ms k l}n}+\frac{\sqrt
    n}{2\iu}\,\frac{\e^{-\frac{k\pi\iu}n}}{\sin(\frac{k\pi}n)}\,.
\end{equation}
The first term in the previous formula can be expressed as $n^{-\frac12}\sum_{j=1}^nf(j)s_j$,
with
\[
f(j)=\sum_{l=j}^n\e^{-\frac{2\pi\iu\ms k l}n}
=\frac{\e^{\frac{k\pi\iu}n}}{2\iu\ms\sin(\frac{k\pi}n)}\,\big(
\e^{-\frac{2\pi\iu\ms k j}n}-\e^{-\frac{2\pi\iu k}n}\big)\,.
\]
Since  $\sum_{j=1}^ns_j=n$, the previous equality
and Eq.~\eqref{deksj} yield the exact identity
\begin{equation}
  \label{idps}
  P(k)=\frac{|\hat s_k|^2}{4\sin^2(\frac{k\pi}n)}\,.
\end{equation}
It should be emphasized that Eq.~\eqref{idps} is in fact valid for {\em any} finite spectrum.
The latter equation shall be the starting point to derive an analytic approximation to the power
spectrum of an energy spectrum satisfying conditions (i) and (ii) above.

In practice, to smooth out spurious fluctuations one usually divides the whole spectrum into
several subspectra with an equal number of levels, and determines the average $\langle
P(k)\rangle$ of the individual power spectra. For this reason, we shall consider a subspectrum
with energies $E_j=E_1+(j-1)d$, where $j=l_0,\dots,l_1\equiv l_0+m$ and $m\gg1$. Calling
$\nu=m/(\ep_{l_1}-\ep_{l_0})$, the spacings $s_i=\nu(\ep_{l_0+i}-\ep_{l_0+i-1})$, $i=1,\dots, m$,
are approximately given by
\[
s_i\simeq \nu F'(E_{l_0+i-1})d=
\frac{\nu d}{\sqrt{2\pi}\si}\,\e^{-\frac{(E_{l_0+i-1}-\mu)^2}{2\si^2}}\,.
\]
The discrete Fourier transform of these spacings is then
\[
\hat s_k\simeq\frac{\nu d}{\sqrt{2\pi
    m}\,\si}\sum_{l=1}^m\e^{-\frac{(E_{l_0}-\mu+(l-1)d)^2}{2\si^2}}\,\e^{-\frac{2\pi\iu\ms kl}{m}}\,.
\]
When $k\ll m$, the previous sum can be approximated with great accuracy by an integral, so that
\[
\hat s_k \simeq \frac{\nu}{\sqrt{2\pi m}\,\si}\,\e^{\frac{2\pi\iu\ms
    k}{md}\,(E_{l_0}-\mu-d)}\,\int_{E_{l_0}-\mu}^{E_{l_1}-\mu}\e^{-\frac{x^2}{2\si^2}}
\e^{-\frac{2\pi\iu\ms k\ms x}{md}}\,\diff x\,.
\]
Setting $x_j=(E_{l_j}-\mu)/(\sqrt2\,\si)$, $y = \sqrt 2\,\pi\si k/(md)$, $z_j=x_j+\iu\ms y$, with
$j=0,1$, and using~\cite[Eq.~7.4.32]{AS70} and Eq.~\eqref{idps}, we finally obtain
\begin{equation}\label{Pksp}
  P(k)\simeq
  \frac{m\,\e^{-2y^2}}{4\sin^2(\frac{\pi k}m)}
  \,\bigg|\frac{\erf z_0-\erf z_1}{\erf x_0-\erf x_1}\bigg|^2\,.
\end{equation}
If one is dealing with the whole spectrum, then $m=n$, $\erf x_1\simeq1$, $\erf x_0\simeq-1$ and
Eq.~\eqref{Pksp} simplifies to
\begin{equation}
  \label{Pkspfull}
  P(k)\simeq
  \frac{n\,\e^{-2y^2}}{16\sin^2(\frac{\pi k}n)}
  \,\big|\erf z_0-\erf z_1\big|^2\,.
\end{equation}
We shall next use Eq.~\eqref{Pksp} to determine the asymptotic behavior of
$P(k)$ in the range $k_0\ll k\ll m$, where $k_0\equiv(md/\si)\max(1,|x_0|,|x_1|)\ll m$
by the condition $\si\gg\sqrt n\,d$. Since $k\ll m$, we can write~\eqref{Pksp} 
as
\begin{equation}
  \label{Pslin}
  P(k)\simeq \frac{m^3\vp(k)}{4\pi^2k^2(\erf x_0-\erf x_1)^2}\,,
\end{equation}
where $\vp(k)\equiv\e^{-2y^2}|\erf z_0-\erf z_1|^2$. The condition $k\gg k_0$ implies that
\begin{equation}
  \label{kcond}
  |z_j|>y\gg 1\,,\quad y\gg |x_j|\,,\qquad j=0,1\,,
\end{equation}
so that the well-known asymptotic formula~\cite[7.1.23]{AS70}
\[
1-\erf z\simeq\frac{\e^{-z^2}}{\sqrt\pi\,z}\,,\qquad |z|\gg 1\,,\quad |\arg z|<\frac{3\pi}4
\]
holds for both $z=z_0$ and $z=z_1$. Taking into account that $2y(x_1-x_0)=2\pi k\in\ZZ$ we
easily obtain
\begin{equation}\label{vp}
\vp(k)\simeq\frac1\pi\,\bigg|\frac{\e^{-x_0^2}}{z_0}-\frac{\e^{-x_1^2}}{z_1}\bigg|^2\,.
\end{equation}
If $x_0^2\simeq x_1^2$, which is only possible when $x_1\simeq-x_0>0$, i.e., when the subspectrum
under consideration is approximately symmetric with respect to $\mu$, Eq.~\eqref{vp} immediately
yields
\[
\vp(k)\simeq
\frac{\e^{-2x_1^2}}{\pi}\,\frac{(x_0-x_1)^2}{|z_0z_1|^2}
\simeq\frac{4x_1^2\,\e^{-2x_1^2}}{\pi y^4}\,.
\]
Substituting into Eq.~\eqref{Pslin} we finally obtain the asymptotic power law
\begin{equation}
  \label{asympeq}
  P(k)\simeq\frac{m^7d^4x_1^2\,\e^{-2x_1^2}}{16\pi^7\si^4(\erf x_1)^2}\,\frac1{k^6}\,.
\end{equation}
In particular, if we are dealing with the full spectrum we have the slightly simpler relation
\begin{equation}
  \label{asympeqfull}
  P(k)\simeq\frac{n^7d^4x_1^2\,\e^{-2x_1^2}}{16\pi^7\si^4}\,\frac1{k^6}\,.
\end{equation}
In the generic case, i.e., when $x_0^2\not\simeq x_1^2$, the leading term of the expansion of
Eq.~\eqref{vp} in powers of $1/y$ becomes
\[
\vp(k)\simeq\frac{\big(\e^{-x_0^2}-\e^{-x_1^2}\big)^2}{\pi\,y^2}\,.
\]
Substituting again into Eq.~\eqref{Pslin} we obtain
\begin{equation}
  \label{asympeqneq}
  P(k)\simeq
  \frac{m^5d^2\big(\e^{-x_0^2}-\e^{-x_1^2}\big)^2}{8\pi^5\si^2(\erf x_0-\erf x_1)^2}\,\frac1{k^4}\,.
\end{equation}
When studying the whole spectrum the previous formula simplifies to
\begin{equation}
  \label{asympeqfullneq}
  P(k)\simeq\frac{n^5d^2\big(\e^{-x_0^2}-\e^{-x_1^2}\big)^2}{32\pi^5\si^2}\,\frac1{k^4}\,.
\end{equation}

Equation~\eqref{asympeqneq} is the key analytic result for establishing the $1/f^4$ behavior of
the averaged power spectrum $\langle P(k)\rangle$ for spin chains of HS type. Indeed, when one
averages the individual $P(k)$ of a large number of subspectra, at most one of these subspectra
can be symmetric about $\mu$. Hence, if the total number of subspectra is sufficiently large we
must have
\begin{equation}\label{avPk}
\langle P(k)\rangle\simeq\frac{m^5d^2}{8\pi^5\si^2}\,
\bigg\langle\bigg(\frac{\e^{-x_0^2}-\e^{-x_1^2}}
{\erf x_0-\erf x_1}\bigg)^{\!\!2}\ms\bigg\rangle\,\frac1{k^4}\,,
\end{equation}
where $m$ is the number of spacings in each subspectrum.

\begin{figure}[h]
  \centering
\includegraphics[draft=false,width=8cm]{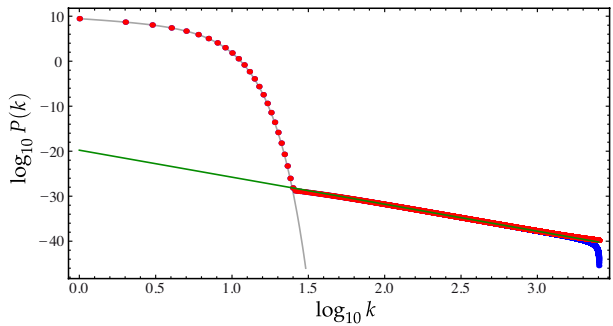}
\caption{$\log$-$\log$ plot of $P(k)$ for the whole spectrum of the spin
  chain~\eqref{cHep} with 100 sites (blue), compared to its analytic
  approximation~\eqref{Pkspfull} (red), the power law~\eqref{asympeqfull} (green line),
  and the approximation~\eqref{Pkspsmall} (gray curve). Note that the blue and red dots are
  practically indistinguishable for $\log_{10}k\lesssim 3$.}
\label{fig:PFB100all}
\end{figure}

In order to test our theoretical results, we shall first consider the ferromagnetic rational
chain of type B and spin $1/2$ introduced in~\cite{YT96}, whose Hamiltonian is given by
\begin{equation}\label{cHep}
\cH^\ep=\sum_{1\le i\neq j\le N}\bigg[
\frac{1-S_{ij}}{(\xi_i-\xi_j)^2}+\frac{1-S_iS_jS_{ij}}{(\xi_i+\xi_j)^2}\bigg]
+\be\sum_{i=1}^N\frac{1-\ep S_i}{\xi_i^2}.
\end{equation}
Here $\be>0$, $\ep=\pm1$, $S_{ij}$ is the operator which permutes the $i$-th and $j$-th spins, $S_i$
is the operator reversing the $i$-th spin, and $\xi_i=\sqrt{2t_i}$, where $t_i$ is the $i$-th zero
of the generalized Laguerre polynomial $L_N^{\be-1}$. It has recently been shown~\cite{BFGR08}
that the partition function of this model is simply given by
\begin{equation}\label{ZPF}
Z(q)=\prod_{j=1}^N\big(1+q^j\big)\,,\qquad q\equiv\e^{-1/(k_{\mathrm{B}}T)}\,,
\end{equation}
so that condition (i) above is satisfied with
\begin{equation}
  \label{E1den}
  E_1=0,\qquad d=1\,,\qquad n=\frac12\,N(N+1)\,.
\end{equation}
It was rigorously proved in Ref.~\cite{EFG09} that when the number of sites $N$ tends to infinity
the level density tends to a Gaussian with parameters
\begin{equation}
  \mu=\frac N4(N+1),\qquad\si^2=\frac{N}{12}(N+\tfrac12)(N+1).
  \label{musi}
\end{equation}
The spectrum is obviously symmetric about its mean, so that if we consider the whole spectrum
$P(k)$ should approximately be given by~\eqref{Pkspfull} for all $k\ll N^2$, and follow the
power law~\eqref{asympeqfull} when $N\ll k\ll N^2$. We have verified that this
prediction is in excellent agreement with the numerical data for a wide range of values of $N$, up
to $N=100$. For instance, in the latter case it is apparent from Fig.~\ref{fig:PFB100all} that the
approximation~\eqref{asympeqfull} is extremely accurate even for $k$ close to $[n/2]$. As to the
$1/f^6$ power law, the relative root mean square (RMS) error in the fit of the $\log_{10}$ of the
RHS of Eq.~\eqref{asympeqfull} to $\log_{10}P(k)$ in the range $50\le k\le 1000\simeq n/5$ is only
$6.9\times 10^{-4}$. The behavior of $P(k)$ for small $k$ seen in Fig.~\ref{fig:PFB100all} perhaps
deserves a brief explanation. To this end, note that when
$k\ll(nd/\si)\min(|x_0|,|x_1|)$, which in this case is tantamount to $k\ll k_0=O(N)$, instead
of~\eqref{kcond} we have $y\ll\min(|x_0|,|x_1|)$. Since, by Eqs.~\eqref{E1den} and \eqref{musi},
$|x_0|=-x_0$ and $|x_1|=x_1$ are $O(N^{1/2})$, when $N$ is large enough both $-z_0$ and $z_1$ are
close to the positive real axis and their modules are at least $O(N^{1/2})$, so that
$\erf(-z_0),\erf{z_1}\simeq1$. From Eq.~\eqref{Pkspfull} we then obtain
\begin{equation}
  \label{Pkspsmall}
  P(k)\simeq
  \frac{n\,\e^{-2y^2}}{4\sin^2(\frac{\pi k}n)}\,,\qquad k\ll k_0\,.
\end{equation}
This is indeed an excellent approximation to $P(k)$ for small $k$, as can be seen from
Fig.~\ref{fig:PFB100all}.

\begin{figure}[h]
  \centering
\includegraphics[draft=false,width=8cm]{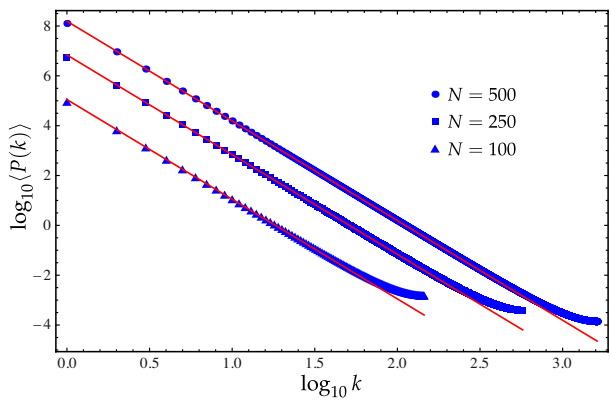}
\caption{$\log$-$\log$ plot of $\langle P(k)\rangle$ for the spin chain~\eqref{cHep} (blue marks),
  compared to its analytic approximation~\eqref{avPk} (red lines). Note that the initial (curved)
  part of the plot of $P(k)$ observed in Fig.~\ref{fig:PFB100all} is absent in this case, since
  $(md/\si)\min(|x_0|,|x_1|)\lesssim 1$ for all subspectra involved.}
\label{fig:PFB100250500marks}
\end{figure}

Even though the spectrum of the chain~\eqref{cHep} is symmetric about $\mu$, the $1/f^4$ power law
predicted for spin chains of HS type clearly emerges when one considers the averaged power spectrum
$\langle P(k)\rangle$ of even a relatively small number of subspectra. To be more precise, we have
divided the energy range $[\mu-5\si,\mu+5\si]$ into ten subspectra of equal length $m=[\si]$ and
computed numerically the resulting $\langle P(k)\rangle$ for a wide range of values of $N$, up to
$N=500$. As can be seen in Fig.~\ref{fig:PFB100250500marks}, Eq.~\eqref{avPk} is in excellent
agreement with the numerical data except for $k$ close to $[m/2]$ in the logarithmic scale. The
accuracy of this approximation steadily improves as $N$ increases; for instance, the relative RMS
error of the approximation~\eqref{avPk} in a $\log$-$\log$ plot for $10\le k\le [m/6]$ and
$N=100$, $250$, and $500$ is respectively given by $0.065$, $0.046$, and $0.033$.

\begin{figure}[h]
  \centering
\includegraphics[draft=false,width=8cm]{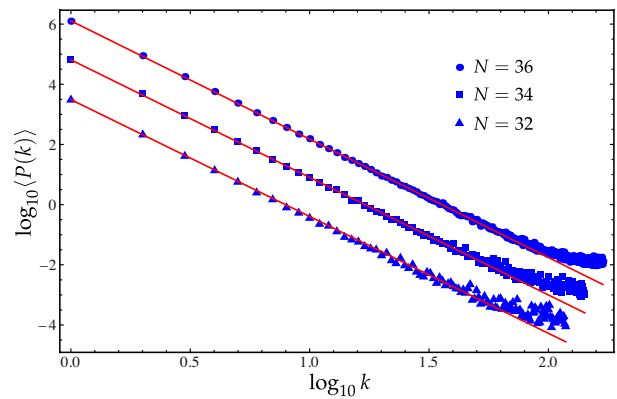}
\caption{Plot of $\log_{10}\langle P(k)\rangle$ for
  the spin chain~\eqref{HS}  (blue marks) fitted
  by a straight line in the range $0\le\log_{10}k\le\log_{10}[m/6]$ (red). The plots
  for $N=32$ and $N=36$ have been displaced to avoid overlapping.}
\label{fig:HSA323436marks}
\end{figure}

Consider next the spin $1/2$ (antiferromagnetic) Haldane--Shastry chain, with
Hamiltonian
\begin{equation}\label{HS}
\cH=\frac12\sum_{1\le i<j\le N}\frac{1+S_{ij}}{\sin^2(\te_i-\te_j)}\,,\qquad \te_j=\frac{j\pi}N\,.
\end{equation}
Although the partition function of this chain has been evaluated in closed form~\cite{FG05}, in
practice its complicated structure precludes the calculation of the spectrum for $N\gtrsim 40$.
On the other hand, the mean and variance of the energy can be exactly computed for all $N$ by
taking suitable traces in Eq.~\eqref{HS}, with the result~\cite{FG05}
\begin{equation}\label{musiHS}
\mu=\frac N{8}(N^2-1)\,,\quad
\si^2=\frac N{480}(N^2-1)(N^2+11)\,.
\end{equation}
It was also numerically shown in the latter reference that for sufficiently large $N$ the level
density is well approximated by a normal distribution with the above parameters. As is well known,
the spectrum of the Haldane--Shastry chain~\eqref{HS} is not equispaced. However, for large $N$
condition (i) above is approximately satisfied, since the vast majority of the differences
$d_i\equiv E_{i+1}-E_i$ are equal to $d=1$ (for even $N$) or $d=2$ (for odd $N$), and the
differences $d_i\ne d$ actually correspond to energies $E_i$ in the tail of the Gaussian level
density; cf.~Ref.~\cite{BFGR08epl}. As to condition (ii), the requirement $\si\gg\sqrt n\,d$ is
also fulfilled for large $N$, since $\si=O(N^{5/2})$, while it was known in~\cite{BFGR08epl} that
$n\le E_{n+1}-E_1=O(N^3)$. For these reasons, it is to be expected that the power spectrum of the
Haldane--Shastry chain obey the $1/f^4$ power law when the number of sites is sufficiently large.
We have numerically verified that this law clearly holds when $N$ ranges from $26$ to $36$.
For instance, in Fig.~\ref{fig:HSA323436marks} we present the results for $N=32,34,36$, where
$\langle P(k)\rangle$ was computed by averaging over ten subspectra of equal maximal length $m$ in
the interval $[\mu-5\si,\mu+5\si]$. A least squares fit of $\log_{10}\langle P(k)\rangle$ to
$\be-\al\log_{10}k$ in the range $1\le k\le[m/6]$ yields an optimum $\al$ of $3.884$, $3.908$, and
$3.933$ for $N=32$, $34$, and $36$, respectively. The corresponding values of the squared
correlation coefficient $r^2$ are $0.9975$, $0.9989$, and $0.9996$, which strongly suggests that
the $1/f^4$ power law also holds in this case when $N$ is sufficiently large.

In conclusion, the results of this letter show that there is a whole family of integrable systems,
namely spin chains of Haldane--Shastry type, the fluctuations of whose spectrum clearly exhibit
$1/f^4$ black noise rather than the $1/f^2$ noise conjectured in Ref.~\cite{FGMMRR04}. Note,
however, that our findings do not invalidate the theoretical justification for the $1/f^2$ law
proposed in the latter reference. Indeed, an essential assumption of this justification is that
the spacings distribution be Poissonian, which is certainly not the case for spin chains of HS
type~\cite{FG05,BB06,BFGR08,BFGR08epl,BFGR09-3}. Given the fact that the spectrum of a
chaotic system features $1/f^\al$ noise with $1\le\al<2$, and that spin chains of HS type possess a
higher degree of integrability than generic integrable systems due to their underlying Yangian
symmetry, it is tempting to conclude that the exponent $\al$ in the $1/f^\al$ spectral noise
provides a quantitative measure of the degree of integrability of a system. It would be of
considerable interest in this respect to find integrable quantum systems featuring $1/f^\al$ noise
with $2<\al\le 4$. Finally, a noteworthy by-product of our analysis is the universal
identity~\eqref{idps}, which shows that $P(k)\propto |\hat s_k|^2/k^2$ for $k\ll n$. This fact
makes it possible to translate any statement about the behavior of the power spectrum of the
statistic $\de_l$ into a simpler statement on the power spectrum of the spacings $s_i$. {}From this
alternative point of view, generic integrable systems (with Poissonian spacings) are characterized
by white noise in the fluctuations of their spacings $s_i$, while in all chaotic systems the
corresponding noise actually grows with the frequency.

\begin{acknowledgments}
This work was supported in part by the MICINN grant~FIS2008-00209 and the UCM--Banco Santander
grant~GR58/08-910556. J.C.B.\ acknowledges the financial support of the
MICINN through an FPU fellowship.
\end{acknowledgments}


\end{document}